\documentclass[12pt]{article}
\usepackage{graphicx}
\usepackage{natbib} 
\usepackage{url} 

\usepackage{amsmath}
\usepackage{amssymb}
\usepackage{amsfonts}
\usepackage{multirow}
\usepackage{amsthm}

\usepackage{moreverb,url}

\usepackage{graphicx}
\usepackage{amsmath}
\usepackage{times}
\usepackage{bm}
\usepackage{enumerate}
\usepackage{mathtools}
\usepackage{setspace}
\usepackage{array,booktabs}
\usepackage{tikz}
\usepackage{natbib}
\usepackage{mathtools}
\usepackage{chngcntr}
\usepackage{algorithm}
\usepackage[utf8]{inputenc}
\usepackage{xcolor, soul}
\usepackage{hyperref}

\newtheorem{thm}{Theorem}
\newtheorem{prop}{Proposition}

\newtheorem{coro}{Corollary}
\DeclareMathOperator*{\esssup}{ess\,sup}

\DeclareMathOperator*{\argmin}{arg\,min}

\theoremstyle{definition}

\newcommand{\blind}{1}

\begin{document}

\def\spacingset#1{\renewcommand{\baselinestretch}%
{#1}\small\normalsize} \spacingset{1}

\date{}

\if1\blind
{
	\title{\bf Deep Neural Networks Guided Ensemble Learning for Point Estimation}
	\author{
		Tianyu Zhan\thanks{Tianyu Zhan is an employee of AbbVie Inc. Corresponding author email address: \texttt{tianyu.zhan.stats@gmail.com}.} \\
		\footnotesize Data and Statistical Sciences, AbbVie Inc., North Chicago, IL, USA\\
		\\
		Haoda Fu\thanks{Haoda Fu is an employee of Eli Lilly and Company. } \\
		\footnotesize Department of Advanced Analytics and Data Sciences, \\ \footnotesize Eli Lilly and Company, Indianapolis, IN, USA \\
		\\
		Jian Kang\thanks{Jian Kang is Professor in the Department of Biostatistics at the University of Michigan, Ann Arbor.} \\
		\footnotesize Department of Biostatistics, University of Michigan, Ann Arbor, MI, USA}
	\maketitle
} \fi

\if0\blind
{
	\title{\bf Deep Neural Networks Guided Ensemble Learning for Point Estimation}
	\maketitle
} \fi

\bigskip
\begin{abstract}
In modern statistics, interests shift from pursuing the uniformly minimum variance unbiased estimator to reducing mean squared error (MSE) or residual squared error. Shrinkage-based estimation and regression methods offer better prediction accuracy and improved interpretation. However, the characterization of such optimal statistics in terms of minimizing MSE remains open and challenging in many problems, for example, estimating the treatment effect in adaptive clinical trials with pre-planned modifications to design aspects based on accumulated data. From an alternative perspective, we propose a deep neural network based automatic method to construct an improved estimator from existing ones. Theoretical properties are studied to provide guidance on applicability of our estimator to seek potential improvement. Simulation studies demonstrate that the proposed method has considerable finite-sample efficiency gain compared to several common estimators. In the Adaptive COVID-19 Treatment Trial (ACTT) as a motivating example, our ensemble estimator essentially contributes to a more ethical and efficient adaptive clinical trial with fewer patients enrolled. The proposed framework can be generally applied to various statistical problems, and can serve as a reference measure to guide statistical research.
\end{abstract}

\noindent%
\vfill
\noindent%

\newpage
\spacingset{2} 

\section{Introduction}

The uniformly minimum variance unbiased estimator is one of the most fundamental and important estimation methods in classical statistics, but its existence and characterization are usually challenging to investigate when one moves beyond exponential families \citep{lehmann2006theory}. In the past several decades, many shrinkage estimation, regression and variable selection methods were proposed \citep{james1992estimation, efron2012large, tibshirani1996regression, varewyck2014shrinkage}. For instance, the James--Stein estimator dominates the maximum likelihood estimator in terms of expected total squared loss beyond two-dimensional Gaussian models \citep{james1992estimation, efron2012large}; lasso offers a better interpretation and prediction accuracy than the ordinary least squares estimates \citep{tibshirani1996regression}. 

However, it remains open and challenging in many problems to identify or construct such optimal statistics with minimized estimation error. For example in the scale-uniform distribution which does not satisfy the usual differentiability assumptions leading to the Cram\'er--Rao bound \citep{galili2016example}, a direct optimization approach may not be feasible. As compared with traditional clinical trials, adaptive clinical trials, for example, the Adaptive COVID-19 Treatment Trial (ACTT) \citep{nihinterim}, are appealing to accommodate uncertainty with limited knowledge of the treatment profiles by allowing prospectively planned modifications to design aspects based on accumulated unblinded data \citep{bretz2009adaptive, chen2010bayesian, chen2014bayesian}. One is interested in an unbiased or consistent estimator of the underlying treatment effect to have an accurate assessment of the efficacy of the study drug, but traditional estimators are often biased \citep{bretz2009adaptive}. Although several methods \citep{shen2001improved, stallard2008estimation} have been proposed to estimate the bias, its correction in adaptive design is still a less well-studied phenomenon, as acknowledged by the Food and Drug Administration [FDA; \cite{fdaguidance}] and the European Medicines Agency [EMA; \cite{committee2007reflection}]. Moving beyond, the next question is how to identify a more efficient estimator of the treatment effect, which further contributes to a more ethical adaptive clinical trial with fewer patients enrolled. 

In this article, we propose a novel deep neural networks guided ensemble learning framework to construct an improved estimator based on existing ones. {This is motivated by the spirit of ensemble learning, for example, XGBoost} \citep{chen2016xgboost} {and Super Learner} \citep{van2007super}, {to build a prediction model by combining the strengths of a collection of simpler base models.} Deep learning techniques are utilized in the proposed method due to their strong functional representation and scalability to large datasets \citep{yang2007studies, goodfellow2016deep, ioannou2020assessment, liu2020masked}. Our framework automatically leverages machine intelligence to construct a feasible estimator, which can serve as a reference measure to guide future statistical research in a particular problem. Based on the obtained theoretical results, we discuss the conditions for our ensemble estimator to gain efficiency, and further provide a few points to consider in general applications. 

The remainder of this article is organized as follows. We introduce our framework in Section \ref{sec:estimator}, and propose a deep learning based algorithm in Section \ref{sec:dnn_alg}. In Section \ref{sec:point}, we study theoretical properties of this estimator and provide practical guidance on implementation. Two simulation studies are conducted in Section \ref{sec:sim} to show efficiency gain. We apply the proposed method to the Adaptive COVID-19 Treatment Trial in Section \ref{sec:adap} to make it more efficient and ethical. Concluding remarks are provided in Section \ref{sec:discussion}.

\section{An ensemble estimator}
\label{sec:estimator}

Our parameter of interest is $\theta$ under an open and bounded $\Theta  \subseteq \mathbb{R}$. For illustration, $\theta$ is considered as a scalar quantity, but the proposed method can be readily applied to a vector as considered in the regression problem at Section \ref{sec:eg2}. Let $\boldsymbol{x} = (x_1, \cdots, x_n)$ be independent and identically distributed (i.i.d.) random variables given on the probability space $(\Omega_x, \mathcal{A}_x, P_x)$, where $\Omega_x$ is a compact set in $\mathbb{R}$ and $P_x = p(x; \theta, \boldsymbol{\omega})$ is the probability function. The nuisance parameters $\boldsymbol{\omega}$ is of $d-1$ dimension with an open and bounded support $\boldsymbol{\Omega} \subseteq \mathbb{R}^{d-1}$ and $d$ is an integer larger than $1$.

Let $T_1(\boldsymbol{x})$ and $T_2(\boldsymbol{x})$ be two estimators of $\theta$. We construct $U(\boldsymbol{x})$ by a linear combination of them,
\begin{equation}
\label{def_ux_motivate}
U(\boldsymbol{x}; w) = w \times T_1(\boldsymbol{x}) + (1-w) \times T_2(\boldsymbol{x}),
\end{equation}
where $w \in \mathbb{R}$. The optimal weight $w^{\{opt\}}$ is the one that minimizes the MSE of $U(\boldsymbol{x}; w)$ for $w \in \mathbb{R}$,
\begin{eqnarray}
w^{\{opt\}} &=& \mathrm{argmin}_{w \in \mathbb{R}} E \left[\left\{ U(\boldsymbol{x}; w) - \theta \right\}^2 \right]  \nonumber\\
&=& \frac{E \left[\left\{ T_2(\boldsymbol{x}) - T_1(\boldsymbol{x}) \right\} \left\{T_2(\boldsymbol{x})-\theta \right\} \right]}{E \left[\left\{ T_1(\boldsymbol{x}) - T_2(\boldsymbol{x}) \right\}^2 \right]}, \label{w_opt}
\end{eqnarray}
where the expectation $E(\cdot)$ is with respect to $P_x$ without being further specified. The estimator in (\ref{def_ux_motivate}) is unbiased if combined with two unbiased estimators. We investigate the reduction of MSE in the following Proposition \ref{prop_w_opt} with proof in the Appendix A of the Supplementary Materials. For simplicity, ``$(\boldsymbol{x})$'' is removed from the notations of $T_1(\boldsymbol{x})$ and $T_2(\boldsymbol{x})$.
\begin{prop}
	\label{prop_w_opt}
	The MSE reduction $\Lambda(w)$ of estimating $\theta$ by $U\left(\boldsymbol{x}; w^{\{opt\}}\right)$ with $w^{\{opt\}}$ in (\ref{w_opt}) as compared with $U(\boldsymbol{x}; w)$ in (\ref{def_ux_motivate}) is
	\begin{align}
	\Lambda(w) = \frac{\left[ E \left\{ \left( T_2 - T_1 \right) \left(T_2-\theta \right) \right\} - E \left\{ \left( T_1 - T_2 \right)^2 \right\} w \right]^2 }{E \left\{ \left( T_1 - T_2 \right)^2 \right\}}. \label{var_opt_imp}
	\end{align}
\end{prop}
This variance improvement $\Lambda(w)$ is non-negative with $\Lambda\left(w^{\{opt\}}\right) = 0$. In some problems where (\ref{w_opt}) is free from $\theta$ and $\boldsymbol{\omega}$ or can be evaluated in a closed form, the solution of $w^{\{opt\}}$ is straightforward -- for example, on estimating the mean of a normal distribution with known coefficient of variation based on two unbiased estimators from the sample mean and the sample variance \citep{khan2015remark}. In general, $w^{\{opt\}}$ in (\ref{w_opt}) is a function of $\theta$, $\boldsymbol{\omega}$ and sample size $n$, but does not necessarily have an analytic solution. For many problems, we do not have closed forms of the distributions of $T_1$ and $T_2$, and thus the direct computation is not feasible. For some other problems, $T_1$ and $T_2$ themselves do not have closed forms, making the computation even harder. 

While it is usually feasible to empirically estimate $w^{\{opt\}}$ given underlying $\theta$ and $\boldsymbol{\omega}$, our goal is to construct improved statistics by building a prospective mapping function $w^{\{opt\}}$ before observing current data. We further denote $\boldsymbol{\phi} = (\theta, \boldsymbol{\omega})$. In the next Section \ref{sec:dnn_alg}, we introduce our proposed algorithm for approximating $w^{\{opt\}}(\boldsymbol{\phi})$ by deep neural networks. 

In this article, we focus on the setting with two base estimators. The generalization to multiple base estimators is provided in Appendix B of the Supplemental Materials with supporting simulation results in Supplemental Table 8. 

\section{A deep learning guide algorithm to approximate $w^{\{opt\}}$}
\label{sec:dnn_alg}

\subsection{Review on deep neural networks}
\label{sec:dnn_review}


We first provide a short review of deep neural networks. Deep learning is a specific subfield of machine learning with a major application to approximate a function ${y} = w(\boldsymbol{\phi})$ \citep{goodfellow2016deep}. We consider the feedforward neural networks, which define a mapping function ${y} = \widetilde{w}(\boldsymbol{\phi};\boldsymbol{\eta})$ and learn values of $\boldsymbol{\eta}$ that result in the best function approximation. 

Consider a motivating example of a neural network with two hidden layers in Supplemental Figure 1. The input parameter $\boldsymbol{\phi}$ has a dimension $d=2$ on the left, with a scaler output $y$ on the right. {The DNN $\widetilde{w}(\boldsymbol{\phi};\boldsymbol{\eta})$ is formulated as,}
\begin{align}
\widetilde{w}(\boldsymbol{\phi};\boldsymbol{\eta}) & = \widetilde{w}^{(3)} \left( \widetilde{w}^{(2)} \left( \widetilde{w}^{(1)} \left( \boldsymbol{x} \right) \right) \right), \nonumber \\
\widetilde{w}^{(h)}(\boldsymbol{x}) & = g_h\left( \boldsymbol{\eta}_{w, h}^T \boldsymbol{x} + \boldsymbol{\eta}_{b, h}  \right), h = 1, 2, 3\nonumber
\end{align}
{where $\widetilde{w}^{(1)}$ is the function connecting the input layer to the first hidden layer, {$\widetilde{w}^{(2)}$} connects the first to the second hidden layer, and $\widetilde{w}^{(3)}$ connects the second hidden layer to the output layer. For $h = 1, 2, 3$, $\boldsymbol{\eta}_{w, h}$ is the weights, $\boldsymbol{\eta}_{b, h}$ the biases, and $g_h()$ is the activation function. To approximate continuous output $y$, the last layer activation function $g_3()$ can be set as the identity function and inner activation functions $g_1()$ and $g_2()$ can be chosen as non-linear functions, e.g., ReLU function $g(z) = \max{(0, z)}$.} The parameter $\boldsymbol{\eta}$ denotes a stack of the weights and bias parameters in the neural networks with dimension $d_{\eta}$. 

We follow the notations in \cite{anthony2009neural} to characterize the complexity of its structure. There are $6$ computation units from the two hidden layers, a total of $18$ weights parameters, and $7$ bias parameters. Therefore, the dimension of $\boldsymbol{\eta}$ is $d_{\eta} = 25$. We define $n^{(l)}$ as the depth and $n^{(w)}$ as the total number of computation units, weights and bias parameters, where $n^{(l)} = 4$ and $n^{(w)} = 31$ in this illustrative example. 


\subsection{Approximation error bound of deep neural networks}
\label{sec:dnn_arch}

We propose to construct a mapping function $\widetilde{w}: \boldsymbol{\Phi} \rightarrow \mathbb{R}$ to approximate $w^{\{opt\}}$ by deep neural networks, where $\boldsymbol{\phi} \in \boldsymbol{\Phi} \subseteq \mathbb{R}^d$. Before studying the approximation error, we first list the following regularity conditions,
\begin{enumerate}
	\item [A.1] Let $\boldsymbol{\Phi}$ of dimension $d$ be open and bounded, with $\partial \boldsymbol{\Phi}$ of class $C^1$.
	\item [A.2] $E \left\{\left( T_1\right)^2; \boldsymbol{\phi} \right\}$, $E \left\{\left( T_2\right)^2; \boldsymbol{\phi} \right\}$,  $E \left( T_1 T_2 ; \boldsymbol{\phi} \right)$, $E \left(T_1; \boldsymbol{\phi} \right)$ and $E \left(T_2; \boldsymbol{\phi} \right)$ are Lipschitz continuous on $\boldsymbol{\Phi}$ for some constants $c_1$, $c_2$, $c_{12}$, $c_4$ and $c_5$, respectively.
	\item [A.3] $T_1$ and $T_2$ have finite second moments bounded by $b_1$ and $b_2$, respectively, for $\boldsymbol{\phi} \in \boldsymbol{\Phi}$.
	\item [A.4] $ \inf_{\boldsymbol{\phi} \in \boldsymbol{\Phi}} E \left\{\left( T_1 - T_2\right)^2 ; \boldsymbol{\phi} \right\} \geq c_L$, for a positive constant $c_L$.   
\end{enumerate}
{\bf Remarks:} Condition A.1 specifies that the parameter space $\boldsymbol{\Phi}$ is open and bounded with a continuously differentiable boundary \citep{evans2010partial}. Condition A.2 requires that the first and second moments of $T_1$, $T_2$ cannot be too steep. A function $u: U \rightarrow \mathbb{R}$ is Lipschitz continuous on $U$ if $\left| u(x) - u(y) \right| \leq C \left| x-y \right|$ for some constant $C$ and every $x, y \in U$. This condition is weaker than differentiation but stronger than continuity. Consider an example where $\boldsymbol{x}$ of size $n$ follows a normal distribution with mean zero and variance $\sigma^2$, and $T_1$ is the sample mean with $E \left\{\left( T_1\right)^2 \right\} = \sigma^2/n$. It can be shown that $C = 1/n$ satisfies the above definition for every $\sigma^2 \in U \subseteq \mathbb{R}^+$. This condition is usually satisfied by $T_1$ and $T_2$ in common statistical models. These two base statistics are required to have finite second moments in Condition A.3. The fourth condition A.4 requires that the variance of $T_1-T_2$ is lower bounded by a positive constant. A trivial counterexample is that the variance of $T_1-T_2$ becomes zero when $T_1=T_2$. We provide more discussion on how to choose $T_1$ and $T_2$ in practice in Section \ref{sec:point_bias_var}. 

In the following Proposition \ref{theorem_1_DNN}, we show that under those four regularity conditions, there exists a neural network $\widetilde{w}(\boldsymbol{\phi};\boldsymbol{\eta}_0)$ with $\boldsymbol{\eta}_0$ as a stack of weight and bias parameters in DNN with finite $n^{(l)}$ and $n^{(w)}$ that can well approximate $w^{\{opt\}}$ with the uniform maximum error defined by,
\begin{equation}
\label{uniform_error}
\left\Vert w^{\{opt\}} - \widetilde{w}  \right\Vert_{\infty} = \max_{\boldsymbol{\phi} \in \boldsymbol{\Phi}} \left| w^{\{opt\}}(\boldsymbol{\phi}) - \widetilde{w}(\boldsymbol{\phi};\boldsymbol{\eta}_0) \right|.
\end{equation}

\begin{prop}
	\label{theorem_1_DNN}
	Under regularity conditions A.1 - A.4, for a given dimension $d$ and an error tolerance $\epsilon_d \in (0, 1)$, there exists a neural network $\widetilde{w}(\boldsymbol{\phi};\boldsymbol{\eta}_0)$ with underlying $\boldsymbol{\eta}_0$ and ReLU activation function that is capable of expressing $w^{\{opt\}}$ with the uniform maximum error
	\begin{equation*}
	\left\Vert w^{\{opt\}} - \widetilde{w}  \right\Vert_{\infty} \leq \epsilon_d.
	\end{equation*} 
	The neural network has a finite number of layers $n^{(l)}$, finite total number of computation units, weight and bias parameters $n^{(w)}$, which satisfy $n^{(l)} < c(d)\left\{\ln(1/\epsilon_d)+1\right\}, $ and \\ $ n^{(w)} < c(d) \epsilon_d^{-d}\left\{\ln(1/\epsilon_d)+1\right\} $, for some constant $c(d)$ depending on $d$. 
\end{prop}

The proof is provided in the Appendix C of the Supplementary Materials. We first show that the objective function $w^{\{opt\}}$ in (\ref{w_opt}) is Lipschitz continuous for $\boldsymbol{\phi} \in \boldsymbol{\Phi}$ under those four regularity conditions. Therefore, it belongs to a Sobolev space $W^{1, \infty}(\boldsymbol{\Phi})$ with the norm 
\begin{equation}
\label{equ_norm}
\left\Vert w \right\Vert_{W^{1, \infty}(\boldsymbol{\Phi})} = \max_{\boldsymbol{m}: |\boldsymbol{m}|\leq 1} \esssup_{\boldsymbol{\phi} \in \boldsymbol{\Phi}}\left| D^{\boldsymbol{m}} w(\boldsymbol{\phi})\right|,
\end{equation}
where $\boldsymbol{m} = (m_1, ..., m_d) \in \left\{0, 1\right\}^d$, $|\boldsymbol{m}| = \sum_{i=1}^d m_i$, $D^{\boldsymbol{m}}$ is the respective weak derivative, and ``$\esssup$'' is the essential supremum \citep{evans2010partial}. The norm $\left\Vert w \right\Vert_{W^{1, \infty}(\boldsymbol{\Phi})}$ in (\ref{equ_norm}) is denoted as $c_d$. Then we obtain the upper bounds on $n^{(l)}$ and $n^{(w)}$ following related results in \citet{yarotsky2017error}.

\subsection{A deep learning based method}
\label{sec:DNN_alg}

\begin{figure}
	\centering
	\includegraphics[scale=0.8]{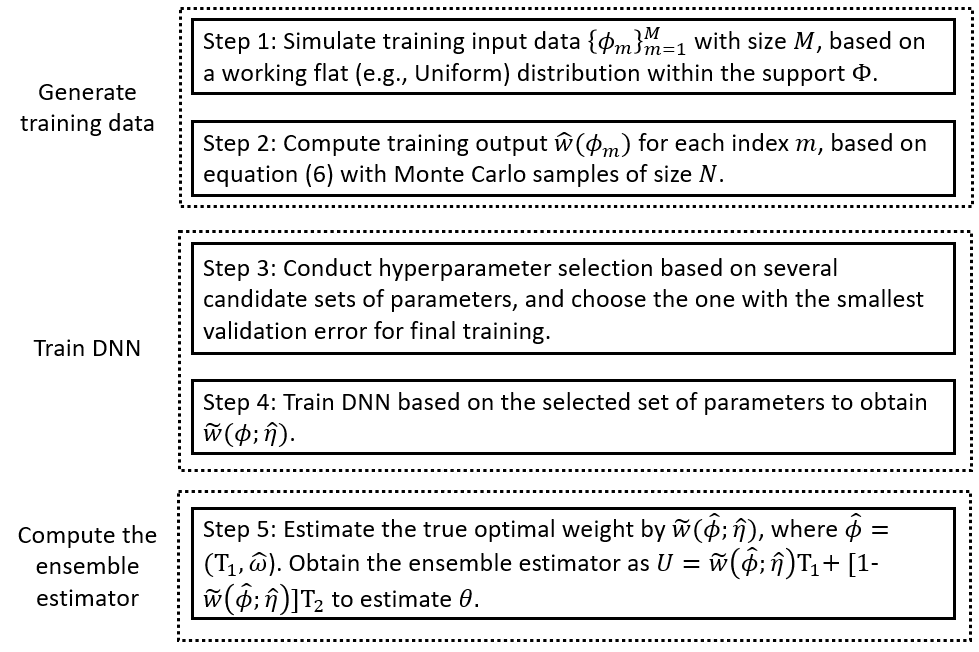}
	\caption{{Diagram of obtaining the ensemble estimator. }}
	\label{F:alg}
\end{figure}

In the previous section, we have shown that there exists a deep neural network $\widetilde{w}(\boldsymbol{\phi};\boldsymbol{\eta}_0)$ that can well approximate $w^{\{opt\}}(\boldsymbol{\phi})$ with a controlled uniform maximum error in (\ref{uniform_error}) by Proposition \ref{theorem_1_DNN}. {Then we follow the diagram in Figure} \ref{F:alg} {to obtain the ensemble estimator. }

At Step 1 and 2 of Figure \ref{F:alg} , we construct input of neural networks as $\left\{\boldsymbol{\phi}_m \right\}_{m=1}^M$, and output as $\left\{\widehat{w}(\boldsymbol{\phi}_m) \right\}_{m=1}^M$ of size $M$. The input $\left\{\boldsymbol{\phi}_m \right\}_{m=1}^M$ are i.i.d. random variables containing $\theta$ and nuisance paramters $\boldsymbol{\omega}$ defined on a working probability space $(\boldsymbol{\Phi}, \boldsymbol{\mathcal{A}}_\phi, \boldsymbol{P}_\phi)$, where $\boldsymbol{\Phi}$ is a compact set in $\mathbb{R}^d$. The working multivariate probability function $\boldsymbol{P}_\phi$ is usually set as some flat distributions to let simulated $\left\{\boldsymbol{\phi}_m \right\}_{m=1}^M$ spread within the support $\boldsymbol{\Phi}$. In the remainder of this article, we draw each of the $d$ elements in $\boldsymbol{\phi}_m$ for $m = 1, \cdots, M$ from $d$ separate uniform distributions under its corresponding support in $\boldsymbol{\Phi}$. The output is $\widehat{w}(\boldsymbol{\phi}_m)$ as an estimate of the underlying $w^{\{opt\}}(\boldsymbol{\phi}_m)$ in (\ref{w_opt}), whose functional form is usually unknown. It can be obtained from the numerical integration method if the joint distribution of $T_1$ and $T_2$ is known, or it can be estimated by the sparse grid method in a high-dimensional setting \citep{shen2010efficient} or by Monte Carlo samples. For a general demonstration, we obtain $\widehat{w}(\boldsymbol{\phi}_m)$ with
\begin{equation}
\label{equ:label_monte}
\widehat{w}(\boldsymbol{\phi}_m) = \frac{\sum_{i=1}^N \big[\left\{ T_2(\boldsymbol{x}_i) - T_1(\boldsymbol{x}_i) \right\} \left\{T_2(\boldsymbol{x}_i) - \theta_m \right\}  \big]}{\sum_{i=1}^N \big[\left\{ T_1(\boldsymbol{x}_i) - T_2(\boldsymbol{x}_i) \right\}^2 \big]},
\end{equation}
where $\boldsymbol{x}_i $ of size $n$ are drawn from the distribution function $p(x; \boldsymbol{\phi}_m)$, for $i=1, \cdots, N$. 

In Step 3 and 4, we train a neural network $\widetilde{w}(\boldsymbol{\phi}_m; \widehat{\boldsymbol{\eta}})$ to approximate $\widehat{w}(\boldsymbol{\phi}_m)$ based on observed data. This can be viewed as a nonlinear regression problem to find the least squared estimator $\widehat{\boldsymbol{\eta}}$ of $\boldsymbol{\eta}_0$ \citep{goodfellow2016deep, bauer2019deep} based on mean squared error loss function, where $\widehat{\boldsymbol{\eta}}$ is given by
\begin{equation}
\label{equ:eta_hat_general}
\widehat{{\boldsymbol{\eta}}} = \argmin_{\boldsymbol{\eta} \in \boldsymbol{H}} \frac{1}{M} \sum_{m=1}^M \left\{ \widehat{w}(\boldsymbol{\phi}_m)- \widetilde{w}(\boldsymbol{\phi}_m; {\boldsymbol{\eta}}) \right\}^2,
\end{equation}
and $\boldsymbol{H}$ is a compact subset of $\mathbb{R}^{d_{\eta}}$; and recall that $d_\eta$ is the dimension of $\boldsymbol{\eta}$. For a generic conclusion, we consider that the $L_2$ error of $\widetilde{w}(\boldsymbol{\phi}; \widehat{\boldsymbol{\eta}})$ is upper bounded by $\epsilon_w$,
\begin{equation}
\label{thm_2_rate_2_general}
E\left\{ \widetilde{w}(\boldsymbol{\phi}; {\boldsymbol{\eta}}_0)- \widetilde{w}(\boldsymbol{\phi}; \widehat{\boldsymbol{\eta}}) \right\}^2 \leq \epsilon_w.
\end{equation}
Under several conditions including bounded empirical risk, Schmidt-Hieber (2020) investigated the convergence rate with $M$ based on certain smoothness indices using a fully connected network with ReLU activation function \citep{schmidt2020nonparametric}. Since $M$ is our design parameter, it can be chosen sufficiently large to control the estimation error $\epsilon_w$. The optimization error of obtaining $\widehat{{\boldsymbol{\eta}}}$ by minimizing non-convex loss function can also be incorporated to $\epsilon_w$ \citep{goodfellow2016deep,bach2017breaking}. {In this article, we adopt the squared loss as a common choice for prediction. Additionally, its scale in squared term is consistent with our objective of minimizing the MSE of the ensemble estimator. Empirical results in Supplemental Materials Table 7 show that the proposed method and the plug-in method have very similar performance, implying that DNN with squared loss can well approximate the true optional weight $w^{\{opt\}}(\boldsymbol{\phi})$.} 


It is important to select a proper network structure by cross-validation at Step 3 \citep{goodfellow2016deep}. We use $80\%$ of data as training data, and the remaining $20\%$ for validation. By increasing the number of layers and number of nodes, the training MSE usually decreases by containing more complex structures. However, the validation error may increase with poor performance at generalization tasks. One can further implement regulation approaches, for example, dropout techniques \citep{goodfellow2016deep}, on the over-saturated structure to decrease validation error below a certain tolerance. {After obtaining this initial working DNN structure, we propose several varying candidate structures around it.} The structure with the smallest validation error within this candidate pool is used in Step 4.


\section{Point estimation of $\theta$}
\label{sec:point}

\subsection{Construct the ensemble estimator}
\label{sec:point_est}

After obtaining $\widetilde{w}(\boldsymbol{\phi}; \boldsymbol{\widehat{\eta}})$ as an estimate of $w^{\{opt\}}(\boldsymbol{\phi})$, we are now ready to construct the ensemble estimator in Figure \ref{F:alg} Step 5. We denote the MSE of $T_1$ and $T_2$ in (\ref{def_ux_motivate}) as $V_1$ and $V_2$, respectively. Suppose that $V_1$ and $V_2$ can be decomposed as follows,
\begin{align}
V_1 & = c^{(n)}_r(\theta, \boldsymbol{\omega}) \times n^{-r}, \label{eq:v1} \\
V_2 & = c^{(n)}_t(\theta, \boldsymbol{\omega}) \times  n^{-t}, \label{eq:v2}
\end{align}
where $r$ and $t$ are positive constants, and the leading terms $c^{(n)}_r(\theta, \boldsymbol{\omega})$ and $c^{(n)}_t(\theta, \boldsymbol{\omega})$ are positive as well. For example, if $T_1$ is the sample mean of $\boldsymbol{x}$ drawn from a Normal distribution with mean $\mu$ and variance $\sigma^2$, then $V_1 = \sigma^2/n$ with $ c^{(n)}_r(\mu, \sigma) = \sigma^2$ and $r = 1$. Without loss of generality, we assume that $V_1 \leq V_2$, which means that $T_1$ is more precise than $T_2$.

Denote $\widehat{\boldsymbol{\omega}}(\boldsymbol{x})$ as an estimator of the nuisance parameters ${\boldsymbol{\omega}}$. Given observed data $\boldsymbol{x}$, we can use $\widehat{\boldsymbol{\phi}} = (T_1, \widehat{\boldsymbol{\omega}})$ to estimate $\boldsymbol{\phi} = (\theta, \boldsymbol{\omega})$, and therefore $\widetilde{w}\left(\widehat{\boldsymbol{\phi}}; \boldsymbol{\widehat{\eta}}\right)$ approximates  $w^{\{opt\}}(\boldsymbol{\phi})$. We plug $\widetilde{w}\left(\widehat{\boldsymbol{\phi}}; \boldsymbol{\widehat{\eta}}\right)$ to equation (\ref{def_ux_motivate}), and compute the ensemble estimator of $\theta$ as $U\left\{\boldsymbol{x}; \widetilde{w}\left(\widehat{\boldsymbol{\phi}}; \boldsymbol{\widehat{\eta}}\right)\right\}$, which is denoted as $U$ for simplicity. 

{Our newly constructed estimator $U$ is a function of data $\boldsymbol{x}$, and does not require further computation after observing $\boldsymbol{x}$. An alternative plug-in approach is to directly estimate $w^{\{opt\}}(\boldsymbol{\phi})$ based on Monte Carlo samples with $\boldsymbol{\phi}$ estimated by $\widehat{\boldsymbol{\phi}}$. However, such a method requires simulations after obtaining data $\boldsymbol{x}$. On the other hand, our method is prospective in the sense that all training processes are done in advance. In the example of adaptive clinical trials of Section} \ref{sec:adap}, {our pre-specified statistic $U$ is more appealing to regulatory agencies to ensure study integrity. Additionally, our proposed method requires a moderate time to train DNN but can instantly calculate $U$ based on well-trained DNN. Section} \ref{sec:adap} {provides more comparisons between the two methods on efficiency and computational time.} 


\subsection{Mean squared error}
\label{sec:point_bias_var}

Before discussing about the MSE of $U$, we introduce two more conditions,
\begin{enumerate}
	\item [B.1] There exists an estimator $\widehat{\boldsymbol{\omega}}(\boldsymbol{x})$ of the nuisance parameter $\boldsymbol{\omega}$, such that $\widetilde{V}_4$ as the maximum of the fourth central moment of $\widehat{\boldsymbol{\omega}}(\boldsymbol{x})$ estimating $\boldsymbol{\omega}$, $T_1$ and $T_2$ estimating $\theta$ is finite. 
	\item [B.2] The maximum of element-wise absolute value of the first order partial derivative \\ $\max_{i = 1, ..., d} \lvert {\partial \widetilde{w} (\boldsymbol{\phi}; {\boldsymbol{\eta}})}/{\partial \boldsymbol{\phi}_i} \rvert$ and $ \max_{i = 1, ..., d_\eta} \lvert {\partial \widetilde{w} (\boldsymbol{\phi}; {\boldsymbol{\eta}})}/{\partial \boldsymbol{\eta}}_i \rvert$ are upper bounded at $\widetilde{c}_{\phi}$ and $\widetilde{c}_{\eta}$, respectively, for $\boldsymbol{\phi} \in \boldsymbol{\Phi}$, $\boldsymbol{\eta} \in \boldsymbol{H}$, $\boldsymbol{\phi}_i$ and $\boldsymbol{\eta}_i$ are the $i$th element of $\boldsymbol{\phi}$ and $\boldsymbol{\eta}$, respectively. 
\end{enumerate}

Condition B.1 requires that the fourth central moment of $\widehat{\boldsymbol{\omega}}(\boldsymbol{x})$, $T_1$ and $T_2$ are upper bounded. This condition is usually satisfied when $V_1$ in (\ref{eq:v1}) and $V_2$ in (\ref{eq:v2}) shrink with increased $n$ and $\widehat{\boldsymbol{\omega}}(\boldsymbol{x})$ is a consistent estimator of $\boldsymbol{\omega}$. {Condition B.2 can be checked empirically based on the fitted neural network $\widetilde{w}(\boldsymbol{\phi}; \boldsymbol{\widehat{\eta}})$. {Specifically, $\max_{i = 1, ..., d_\eta} \lvert {\partial \widetilde{w} (\boldsymbol{\phi}; {\boldsymbol{\eta}})}/{\partial \boldsymbol{\eta}}_i \rvert$ is usually small because gradient descent type algorithms are used in DNN training to obtain $\widehat{\boldsymbol{\eta}}$, and one can narrow $\boldsymbol{H}$ to make this value bounded.} Next, we provide upper bounds on the mean square error of $U$ in the following Theorem \ref{theorem_3_bias}. 

\begin{thm}
	\label{theorem_3_bias}
	Under the aforementioned conditions A.1 - A.4, B.1, B.2, and (\ref{thm_2_rate_2_general}), the MSE of the ensemble estimator $U$ is upper bounded at
	\begin{equation}
	\label{thm3_mse}
	E\left( U - \theta \right)^2 \leq E\left(\widetilde{U}-\theta \right)^2 + S_1 + S_2,
	\end{equation}
	where
	\begin{align}
	\widetilde{U}  = &  w^{\{opt\}}(\boldsymbol{\phi}) T_1 + \left\{ 1-w^{\{opt\}}(\boldsymbol{\phi}) \right\} T_2, \label{thm3_U_tilde} \\
	S_1 = &  s_1(\widetilde{c}_\theta, d) \widetilde{V}_4 +  s_2(\widetilde{c}_\theta, d) {V}_2^{1/2} \widetilde{V}_4^{1/2}, \label{thm3_S1} \\
	S_2 = & 4 \epsilon_d^2 V_2 + 4 \epsilon_d {V}_2^{1/2} \sqrt{s_1(\widetilde{c}_\theta, d) \widetilde{V}_4 + s_2(\widetilde{c}_\theta, d)\widetilde{V}_4^{1/2} {V}_2^{1/2} + {V}_2}, \label{thm3_S2} 
	\end{align}
	{for some constants $s_1(\widetilde{c}_\theta, d)$ and $s_2(\widetilde{c}_\theta, d)$ depending on $\widetilde{c}_\theta$ and $d$, $\epsilon_d$ from Proposition} \ref{theorem_1_DNN}, {${V}_2$ in }(\ref{eq:v2}) {and $\widetilde{V}_4$ in Condition B.1.} 
\end{thm}

The proof and functional forms of $s_1(\widetilde{c}_\theta, d)$ and $s_2(\widetilde{c}_\theta, d)$ are provided in the Appendix D of the Supplementary Materials. $E\left(\widetilde{U}-\theta \right)^2$ in (\ref{thm3_mse}) is the MSE of $\widetilde{U}$ (\ref{thm3_U_tilde}) with the unknown optimal weight $w^{\{opt\}}$. 


In the next Corollary \ref{cor:improv}, we investigate sufficient conditions for $U$ to have a smaller MSE than $T_1$, which is more precise than $T_2$.
\begin{enumerate}
	\item [C.1] $ {S_1 + S_2} < \left(1- w^{\{opt\}} \right)^2 \left\{\sqrt{var(T_1)}+\sqrt{var(T_2)} \right\}^2$ and $w^{\{opt\}} \neq 1$.
	\item [C.2] $cov(T_1, T_2) < -\left(S_1 + S_2\right)/{\left(1-w^{\{opt\}}\right)^2}/2 + var(T_1)/2 + var(T_2)/2 $.  
\end{enumerate}
\begin{coro}
	\label{cor:improv}
	If conditions A.1 - A.4, B.1, B.2, C.1 and C.2 and (\ref{thm_2_rate_2_general}) are satisfied, then the ensemble estimator $U$ has a smaller MSE than $T_1$. 
\end{coro}

Both conditions C.1 and C.2 can be empirically checked based on specific choices of $\theta$ and $\boldsymbol{\omega}$ in the validation stage with numerical errors acknowledged. One can simulate data from the parametric distribution in Section \ref{sec:estimator} to estimate $\widetilde{V}_4$ and $V_2$ in $S_1$ by their empirical counterparts. Numerical methods can be implemented to estimate $\widetilde{c}_\theta$. $S_2$ is usually negligible in practice due to small $\epsilon_d$ by properly choosing a DNN structure as discussed in Section \ref{sec:DNN_alg}. 

As noted in Corollary 1, C.1 and C.2 are sufficient but not necessary conditions for the improvement of $U$, in the sense that the lower bound of the MSE improvement is positive. When C.1 and C.2 are not satisfied, one can still implement our method to obtain $U$ and evaluate its performance in the validation stage. In particular problems where $w^{\{opt\}}$ is close to $1$, C.1 may not be met because its right-hand side is approximately zero. This finding is also useful to support that the base estimator $T_1$ itself is a feasible choice in practice. Essentially, C.1 makes sure that the upper bound of $cov(T_1, T_2)$ in C.2 is achievable. Guided by C.2, we can choose two estimators that are not highly correlated to seek improvement.

\section{Simulations}
\label{sec:sim}

\subsection{{Scale-uniform} family of distributions}
\label{sec:eg1}

In this section, we consider the {scale-uniform} distribution $Unif\Big([1-k]\theta, [1+k]\theta \Big)$ with the parameter of interest $\theta$ and a known design parameter $k \in (0, 1)$ \citep{galili2016example}, where $Unif$ denotes the Uniform distribution. Let $p_x(x;\theta, k)$ be the probability density function of $Unif\Big([1-k]\theta, [1+k]\theta \Big)$. Since the support $\Omega_x = \left\{ x \in \mathbb{R}: p_x(x;\theta, k)>0 \right\}$ is not the same for all $\theta \in \Theta$ with $\Theta$ as an open interval in $\mathbb{R}$, this distribution family does not satisfy the usual differentiability assumptions leading to the Cram\'er--Rao bound and efficiency of maximum likelihood estimators \citep{lehmann2006theory, galili2016example}. While it can be challenging to optimize MSE directly, we apply the proposed method to construct a more efficient estimator of $\theta$ based on existing ones. 

As a starting point, we utilize the Rao--Blackwell Theorem to construct $T_1$ as an improved unbiased estimator. The minimal sufficient statistic for $\theta$ is $\left\{ x_{(1)}, x_{(n)} \right\}$, where $x_{(1)} = \min(\boldsymbol{x})$ and $x_{(n)} = \max(\boldsymbol{x})$. Since $x_1$ is unbiased for $\theta$, then by the Rao--Blackwell theorem we have,
\begin{equation}
\label{sim1_rb}
T_1 = \widehat{\theta}_{RB} = E\Big[ x_1 | x_{(1)}, x_{(n)} \Big] = \frac{x_{(1)}+x_{(n)}}{2}.
\end{equation}
We consider the James--Stein estimator $\widehat{\theta}_{JS}$ \citep{james1992estimation} as the second estimator $T_2$: 
\begin{equation}
T_2 = \widehat{\theta}_{JS} = \left[1-\left\{(p-2)\widehat{\sigma}^2\right\} \middle/ \left\{pn \left(\widehat{\theta}_{M}\right)^2\right\} \right] \widehat{\theta}_{M},
\end{equation}
where $\widehat{\sigma}^2 = k^2 \left(\widehat{\theta}_{M}\right)^2 / 3$ is an estimator of variance and $\widehat{\theta}_{M} = x_{(n)}/\left\{1+k(n-1)/(n+1)\right\}$ is the unbiased corrected version of the maximum likelihood estimator \citep{galili2016example}. Since $\theta$ is a numerical value in this example, we set the dimension $p=3$ for illustrating purposes. 

We combine $T_1 = \widehat{\theta}_{RB}$ and $T_2 = \widehat{\theta}_{JS}$ in (\ref{def_ux_motivate}) to construct a new estimator $U\left( \widehat{\theta}_{RB}, \widehat{\theta}_{JS}  \right)$ based on the proposed method. Suppose we are interested in $\theta \in \Theta = (0.5, 5)$ as an open interval in $\mathbb{R}$ with finite data size $n = 10$. We simulate $M = 10^3$ training input data of varying $\theta$ from $Unif(0.2, 10)$ with a wider range to cover $(0.5, 5)$ and the known parameter $k$ at either $0.1$ or $0.9$ to accommodate the scenarios considered at Table \ref{sim:eg1} for performance evaluation. The above training data sample spaces can be set wider as needed. {Supplemental Table 2 and 3 evaluate a narrower range of $\theta$ when training DNN, and results are generally robust but slightly worse than Table} \ref{sim:eg1}. {Therefore, it is recommended to set a relatively wide range of $\theta$ in the training stage.}

The input data of the neural network is $\boldsymbol{\phi} = (\theta, k)$, and the output $\widehat{w}(\boldsymbol{\phi})$ in (\ref{equ:label_monte}) is evaluated by $N = 10^6$ Monte Carlo samples. In cross-validation, we consider 4 candidate structures: $2$ hidden layers with $40$ nodes per layer, $2$ hidden layers with $60$ nodes per layer, $3$ hidden layers with $40$ nodes per layer, $3$ hidden layers with $60$ nodes per layer. We use a dropout rate of $0.1$, number of training epochs at $10^3$, and a batch size of $100$ in the training process to obtain a fitted network $\widetilde{w}(\boldsymbol{\phi}; \boldsymbol{\widehat{\eta}})$. The above hyperparameters can be chosen by cross-validation, and are utilized throughout this article if not specified otherwise. The training process is implemented by the R package {\texttt{keras}} \citep{all2020a} with more details in our shared R code. The number of iterations for testing at Table \ref{sim:eg1} is $10^6$. 

To evaluate the efficiency gain of our method, we compute the relative efficiency of $U\left( \widehat{\theta}_{RB}, \widehat{\theta}_{JS}  \right)$ versus five existing estimators: $\widehat{\theta}_{RB}$, $\widehat{\theta}_{JS}$, $\widehat{\theta}_{M}$, $\widehat{\theta}_{RB, JS}=(\widehat{\theta}_{RB}+\widehat{\theta}_{JS})/2$ and $\widehat{\theta}_E$ as the sample mean. The relative efficiency of two estimators is defined as the inverse ratio of their MSEs. Under several scenarios in Table \ref{sim:eg1}, the ensemble estimator $U\left( \widehat{\theta}_{RB}, \widehat{\theta}_{JS} \right)$ is uniformly more efficient than four comparators, as demonstrated by all ratios larger than $1$. We also validate that both conditions C.1 and C.2 in Section \ref{sec:point_bias_var} are satisfied under values of $k$ and $\theta$ in Table \ref{sim:eg1}. These findings further support the observed MSE improvement. Additionally, a minimax study that computes the minimum of relative efficiency gain with respect to $\theta \in (0.5, 5)$ shows consistent findings. We observe that $\widehat{\theta}_{RB}$ is more efficient than $\widehat{\theta}_{JS}$ when $k=0.1$, and vice versa when $k=0.9$. Our $U\left( \widehat{\theta}_{RB}, \widehat{\theta}_{JS} \right)$ learns their advantages under different $k$'s and shows a consistently better performance. 

{On computational time, it takes $58$ minutes to simulate training data with $M = 10^3$, $59$ seconds to conduct hyperparameter tuning with $4$ sets of parameters and less than $10$ seconds to train the final DNN model. With the fitted DNN model, it only takes $9.2$ minutes to generate results in Table} \ref{sim:eg1} {with $10^6$ simulation iterations for each scenario}.   

We further evaluate our proposed method versus directly applying Support Vector Machine (SVM; \citet{boser1992training}), {Gaussian Process Regression (GPR;} \citet{williams2006gaussian}), {XGBoost} \citep{chen2016xgboost} {and Super Learner (SL;} \citet{van2007super}) with base models of SVM, XGBoost and neural networks. Their bootstrap training data size is $1,000$, which is the same as our training data size. The proposed method has a consistently better performance than the four methods with results in Supplemental Table 5. Additional analysis with $n=2$ in the Supplemental Table 1 of the Supplementary Materials also demonstrates the superior performance of the proposed method. 

\begin{table}
	\small
	\caption{High relative efficiency of $U\left( \widehat{\theta}_{RB}, \widehat{\theta}_{JS} \right)$ versus two base components $\widehat{\theta}_{RB}$ and $\widehat{\theta}_{JS}$, $\widehat{\theta}_{RB, JS}$ as a direct average of two base estimators, the bias-corrected maximum likelihood estimator $\widehat{\theta}_{M}$ and the sample mean $\widehat{\theta}_{E}$.}
	\label{sim:eg1}
	\centering
	\begin{tabular}{cccccccccc}
		\multicolumn{2}{c}{} & & \multicolumn{1}{c}{} && \multicolumn{5}{c}{Relative efficiency versus} \\ 
		$k$ & $\theta$ &  & Root of MSE && $\widehat{\theta}_{RB}$& $\widehat{\theta}_{JS}$ & $\widehat{\theta}_{RB, JS}$ & $\widehat{\theta}_{M}$ & $\widehat{\theta}_{E}$ \\ [5pt]
		0.1 & 0.5 &  & 0.006 &  & 1.008 & 1.566 & 1.109 & 1.566 & 2.220 \\
		0.1 & 1 &  & 0.012 &  & 1.009 & 1.564 & 1.108 & 1.564 & 2.221 \\
		0.1 & 5 &  & 0.061 &  & 1.008 & 1.566 & 1.109 & 1.566 & 2.211 \\
		0.1 & (0.5, 5) &  & - &  & 1.008 & 1.564 & 1.108 & 1.564 & 2.211 \\
		\\
		0.9 & 0.5 &  & 0.043 &  & 1.679 & 1.003 & 1.149 & 1.010 & 3.699 \\
		0.9 & 1 &  & 0.086 &  & 1.680 & 1.003 & 1.149 & 1.010 & 3.691 \\
		0.9 & 5 &  & 0.428 &  & 1.676 & 1.003 & 1.148 & 1.010 & 3.690 \\
		0.9 & (0.5, 5) &  & - &  & 1.674 & 1.003 & 1.147 & 1.010 & 3.677 \\
	\end{tabular}
\end{table}

\subsection{Regression model for analyzing heterogeneous data}
\label{sec:eg2}

Aggregating and analyzing heterogeneous data is one of the most fundamental challenges in scientific data analysis \citep{fan2014challenges}. For observable $\boldsymbol{X} \in \mathbb{R}^d$ and a discrete variable $Z \in \mathcal{Z}$, a general mixture model assumes $\boldsymbol{X}|(Z = z) \sim \mathcal{F}(\boldsymbol{\theta}_z) $ for a distribution $\mathcal{F}$ with parameters $\boldsymbol{\theta}_z$ in the sub-population $z$ \citep{fan2018curse}. The variable $Z$ can be known in some applications, for example, synthesizing control information from multiple historical clinical trials \citep{neuenschwander2010summarizing}; or it can be latent in general \citep{fan2014challenges}.  


In this motivating simulation study, we consider the following Gaussian regression model where the variance of the dependent variable is proportional to the square of its expected value \citep{amemiya1973regression},
\begin{equation}
\label{equ:sim2_assump}
y_i \sim \mathcal{N}\left(\boldsymbol{x}_i^\prime \boldsymbol{\theta}, \left[\boldsymbol{x}_i^\prime \boldsymbol{\theta}\right]^2\right),
\end{equation}
where $\boldsymbol{x}_i$ is a vector of covariates for subject $i$, and $\boldsymbol{\theta}$ is a vector of unknown parameters. 

Challenges exist in this problem to find an efficient estimator of $\boldsymbol{\theta}$ in finite samples. The minimal sufficient statistics consisting of sample mean and sample variance are not complete for $\boldsymbol{\theta} \in \Theta$ \citep{khan2015remark}. When $\boldsymbol{x}_i^\prime \boldsymbol{\theta}$ is relatively small, the Fisher information matrix can be ill-conditioned \citep{amemiya1973regression}, which may introduce substantial bias in the maximum likelihood estimator (MLE). As robust alternatives, \cite{amemiya1973regression} considers the following estimators,
\begin{align}
\widehat{\boldsymbol{\theta}}_L & = \left[\sum_{i=1}^n \boldsymbol{x}_i \boldsymbol{x}_i^\prime \right]^{-1} \sum_{i=1}^n \boldsymbol{x}_i y_i \nonumber \\
\widehat{\boldsymbol{\theta}}_W & = \left[\sum_{i=1}^n \frac{1}{\left(\boldsymbol{x}_i^\prime \widehat{\boldsymbol{\theta}}_L \right)^2}\boldsymbol{x}_i \boldsymbol{x}_i^\prime \right]^{-1} \sum_{i=1}^n \frac{1}{\left(\boldsymbol{x}_i^\prime \widehat{\boldsymbol{\theta}}_L \right)^2} \boldsymbol{x}_i y_i  \nonumber,
\end{align}
where $\widehat{\boldsymbol{\theta}}_L$ is the least square estimator and $\widehat{\boldsymbol{\theta}}_W$ is the weighted least square estimators. We upper bound the weight ${1}/{\left(\boldsymbol{x}_i^\prime \widehat{\boldsymbol{\theta}}_L \right)^2}$ by $10^5$ to avoid extreme values. {Quantile regression estimator $\widehat{\boldsymbol{\theta}}_Q$} \citep{koenker2005quantile} {can also be applied to estimate $\boldsymbol{\theta}$ based on heterogeneous data. We utilize our proposed method to assemble $\widehat{\boldsymbol{\theta}}_Q$ as $\boldsymbol{T}_1$ and $\widehat{\boldsymbol{\theta}}_W$ as $\boldsymbol{T}_2$ to obtain a better estimator $\boldsymbol{U}\left(\widehat{\boldsymbol{\theta}}_Q, \widehat{\boldsymbol{\theta}}_W \right)$.} 

In this study, we consider that $\boldsymbol{\theta}$ is a four-dimensional vector with $\theta_1$ as intercept and $\theta_2$, $\theta_3$ and $\theta_4$ as coefficients. The parameter space for $\theta_1$, $\theta_2$, $\theta_3$ and $\theta_4$ are considered at $\Theta = (-1.5, 1.5)$. Covariates $\boldsymbol{x}_i$, for $i = 1, \cdots, n$, are simulated from a uniform distribution with a lower bound $-2$ and an upper bound $2$. A moderate sample size of $n=100$ is evaluated in this study. The number of Monte Carlo samples is $N = 10^5$ when computing $\widehat{w}(\boldsymbol{\phi})$ in (\ref{equ:label_monte}). 

{Table} \ref{sim:eg2_re} {shows that our estimator $\boldsymbol{U}\left(\widehat{\boldsymbol{\theta}}_Q, \widehat{\boldsymbol{\theta}}_W \right)$ is consistently more efficient than three comparators, as demonstrated by relative efficiencies larger than one under all scenarios. Even though $\widehat{\boldsymbol{\theta}}_Q$ is generally more efficient than $\widehat{\boldsymbol{\theta}}_W$, our ensemble estimator $\boldsymbol{U}\left(\widehat{\boldsymbol{\theta}}_Q, \widehat{\boldsymbol{\theta}}_W \right)$ can still decrease MSE based on those two base estimators. }

\begin{table}
	\small
	\caption{{High relative efficiency of $\boldsymbol{U}\left(\widehat{\boldsymbol{\theta}}_Q, \widehat{\boldsymbol{\theta}}_W \right)$ versus two base components $\widehat{\boldsymbol{\theta}}_Q$ and $\widehat{\boldsymbol{\theta}}_W$, the least square estimator $\widehat{\boldsymbol{\theta}}_L$.}}
	\centering
	\begin{tabular}{cccccccc}
		\multicolumn{4}{c}{} && \multicolumn{3}{c}{Relative efficiency versus}  \\ 
		$\theta_1$ & $\theta_2$ & $\theta_3$ & $\theta_4$ & &
		$\widehat{\boldsymbol{\theta}}_Q$ & $\widehat{\boldsymbol{\theta}}_W$ & $\widehat{\boldsymbol{\theta}}_L$   \\ [5pt]
0.2 & 0.2 & 0.2 & -0.2 &  & 1.089 & 1.524 & 1.677 \\
0.2 & 0.2 & -0.2 & -0.2 &  & 1.068 & 1.457 & 1.655 \\
0.2 & -0.2 & -0.2 & -0.2 &  & 1.045 & 1.521 & 1.486 \\
\\
1.2 & 1.2 & 1.2 & -1.2 &  & 1.150 & 1.832 & 1.783 \\
1.2 & 1.2 & -1.2 & -1.2 &  & 1.090 & 1.867 & 1.685 \\
1.2 & -1.2 & -1.2 & -1.2 &  & 1.041 & 1.659 & 1.489 \\
	\end{tabular}
	\label{sim:eg2_re}
\end{table}

\section{Adaptive COVID-19 Treatment Trial}
\label{sec:adap}


In this section, we apply our method to the Adaptive COVID-19 Treatment Trial to evaluate the safety and efficacy of remdesivir from Gilead Inc. in hospitalized adults diagnosed with COVID-19 \citep{nihinterim}. We conduct simulation studies based on summary statistics obtained from the literature. For illustration, we consider the sample size reassessment adaptive design with a binary endpoint of achieving hospital discharge on Day 14.  {The objective is to estimate the treatment effect $\theta = \theta_2- \theta_1$, where $\theta_1$ and $\theta_2$ are the response rates of achieving such binary endpoint in the placebo and the treatment, respectively. The underlying $\theta_1 = 0.47$ and $\theta_2 = 0.59$ are assumed based on interim results} \citep{nihinterim}. {In this case study, we estimate the treatment effect $\theta$ with unknown $\theta_1$ and $\theta_2$.} 

We consider a two-stage adaptive design, where $n^{(1)}$ subjects are randomized to the treatment group and $n^{(1)}$ subjects to the control group in the first stage. After observing interim data from those $2\times n^{(1)}$ subjects, we adjust sample size based on the following rule,
\begin{equation}
\label{adaptive_rule}
n^{(2)} = \left\{
\begin{array}{ll}
n^{(2)}_{min}, \:\:\:\:\:\:\:\:\:\:\:\: \text{if} \:\:\: \widehat{\theta}\left\{\boldsymbol{x}_2^{(1)}\right\} - \widehat{\theta}\left\{\boldsymbol{x}_1^{(1)}\right\} > \theta_{min} \\
n^{(2)}_{max}, \:\:\:\:\:\:\:\:\:\: \text{otherwise}
\end{array}
\right.
\end{equation}
where $\widehat{\theta}\left\{\boldsymbol{x}^{(h)}_j\right\}$ is the sample average, $\boldsymbol{x}^{(h)}_j$ is a vector of observed binary data of size $n^{(h)}$ for group $j$, $j=1, 2$ at stage $h$, $h=1, 2$, and $n^{(2)}_{min}$, $n^{(2)}_{max}$ and $\theta_{min}$ are pre-specified design features. Basically, $n^{(2)}$ in the second stage is decreased to $n^{(2)}_{min}$ if a promising treatment effect larger than a clinically meaningful difference $\theta_{min}$ is observed, but increased to $n^{(2)}_{max}$ otherwise. We apply the proposed method to this problem to construct an improved estimator. 

We consider $\theta_1 \in (0.2, 0.7)$ and $\theta \in (-0.2, 0.3)$ as our parameter spaces, and $n^{(1)} = 100$, $n^{(2)}_{min} = 50$, $n^{(2)}_{max} = 250$ and $\theta_{min} = 0.16$ as design features in (\ref{adaptive_rule}). We first build an unbiased estimator of $\theta$ as the weighted average of treatment effects from two stages \citep{bretz2009adaptive}, 
\begin{equation}
\label{adap_T1}
\widetilde{\theta}(k) = k \Delta^{(1)} + (1-k) \Delta^{(2)}, 
\end{equation} 
where $k \in [0, 1]$ is a constant, and $\Delta^{(h)} = \widehat{\theta}\left\{\boldsymbol{x}^{(h)}_2\right\} - \widehat{\theta}\left\{\boldsymbol{x}^{(h)}_1\right\} $ is based on data at stage $h$, for $h=1, 2$. We combine $T_1 = \widetilde{\theta}(0.5)$ and $T_2 = \Delta^{(1)}$ in our framework to deliver a more accurate estimator within a neighborhood of the underlying $\theta$. The input data for neural network is $\boldsymbol{\phi} = (\theta_1, \theta)$. The nuisance parameter $\theta_1$ is estimated by $0.5\widehat{\theta}\left\{\boldsymbol{x}^{(1)}_1\right\}+0.5\widehat{\theta}\left\{\boldsymbol{x}^{(2)}_1\right\}$. Three comparators are evaluated as well: $\widetilde{\theta}(0.2)$, $\widetilde{\theta}(0.5)$ and $\widetilde{\theta}(0.8)$ in (\ref{adap_T1}) with $k = 0.2, 0.5$, and $0.8$, respectively. 

Our estimator is more efficient than three comparators, supported by the high relative efficiency in Table \ref{sim:table_case}. In particular, it learns the better efficiency of $\widetilde{\theta}(0.2)$ under $\theta=0$ and  $\widetilde{\theta}(0.5)$ under $\theta>0$ to achieve a superior performance under all scenarios evaluated. Additionally, our method also has a small bias by combining the two unbiased estimators. Additional analysis in the Supplemental Table 6 of the Supplementary Materials shows consistent findings. Supplemental Table 8 of the Supplementary Materials demonstrates that our method constructs a more efficient estimator by combining three estimators in adaptive designs with three stages. 

{As compared with the plug-in approach discussed in Section} \ref{sec:point_est}, {our method can obtain the functional form of $U$ before observing data. This pre-specified feature is more appealing to regulatory agencies to ensure study integrity. Supplemental Table 7 shows that both methods have very similar MSE, and Supplemental Materials Section 5 discusses the saving in computational time of our method when there are a large number of simulation iterations in the testing stage. }

In terms of hypothesis testing to detect a promising treatment effect, we plot the power of rejecting the one-sided null hypothesis $H_0: \theta \leq 0$ at a type I error rate $\alpha = 5\%$ under $\theta_1 = 0.47$ and varying treatment effect $\theta$ in Figure \ref{F:power}. The critical values of rejecting $H_0$ are computed at $0.064$ for our method, $0.064$ for $\widetilde{\theta}(0.2)$, $0.068$ for $\widetilde{\theta}(0.5)$, and $0.094$ for $\widetilde{\theta}(0.8)$ by the grid search method to control validating type I error rates not exceeding $5\%$ when $\theta_1 = \theta_2 \in (0.42, 0.66)$. Our proposed method has a consistently higher power of detecting a promising treatment effect than the other three estimators. A more ethical and efficient adaptive clinical trial can be designed based on our proposed method to evaluate treatment options to cure COVID-19. 

\begin{table}
	\small
	\caption{High relative efficiency of $U\left\{ \widetilde{\theta}(0.5), \Delta^{(1)} \right\}$ as compared with three unbiased estimators in the the Adaptive COVID-19 Treatment Trial.}
	\centering
	\begin{tabular}{cccccccccc}
		\multicolumn{3}{c}{} && \multicolumn{2}{c}{$U\left\{ \widetilde{\theta}(0.5), \Delta^{(1)} \right\}$} && \multicolumn{3}{c}{Relative efficiency versus} \\ 
		$\theta_1$ & $\theta_2$ & $\theta$ && Bias & Root of MSE  && $\widetilde{\theta}(0.2)$ & $\widetilde{\theta}(0.5)$ & $\widetilde{\theta}(0.8)$ \\ [5pt]
		0.47 & 0.47 & 0 &  & 0.001 & 0.039 &  & 1.012 & 1.164 & 2.154 \\
		& 0.57 & 0.1 &  & 0.001 & 0.045 &  & 1.195 & 1.035 & 1.621 \\
		& 0.59 & 0.12 &  & 0.001 & 0.047 &  & 1.286 & 1.018 & 1.484 \\
		& 0.61 & 0.14 &  & $< 0.001$ & 0.050 &  & 1.383 & 1.007 & 1.355 \\
	\end{tabular}
	\label{sim:table_case}
\end{table}

\begin{figure}
	\centering
	\includegraphics[scale=0.21]{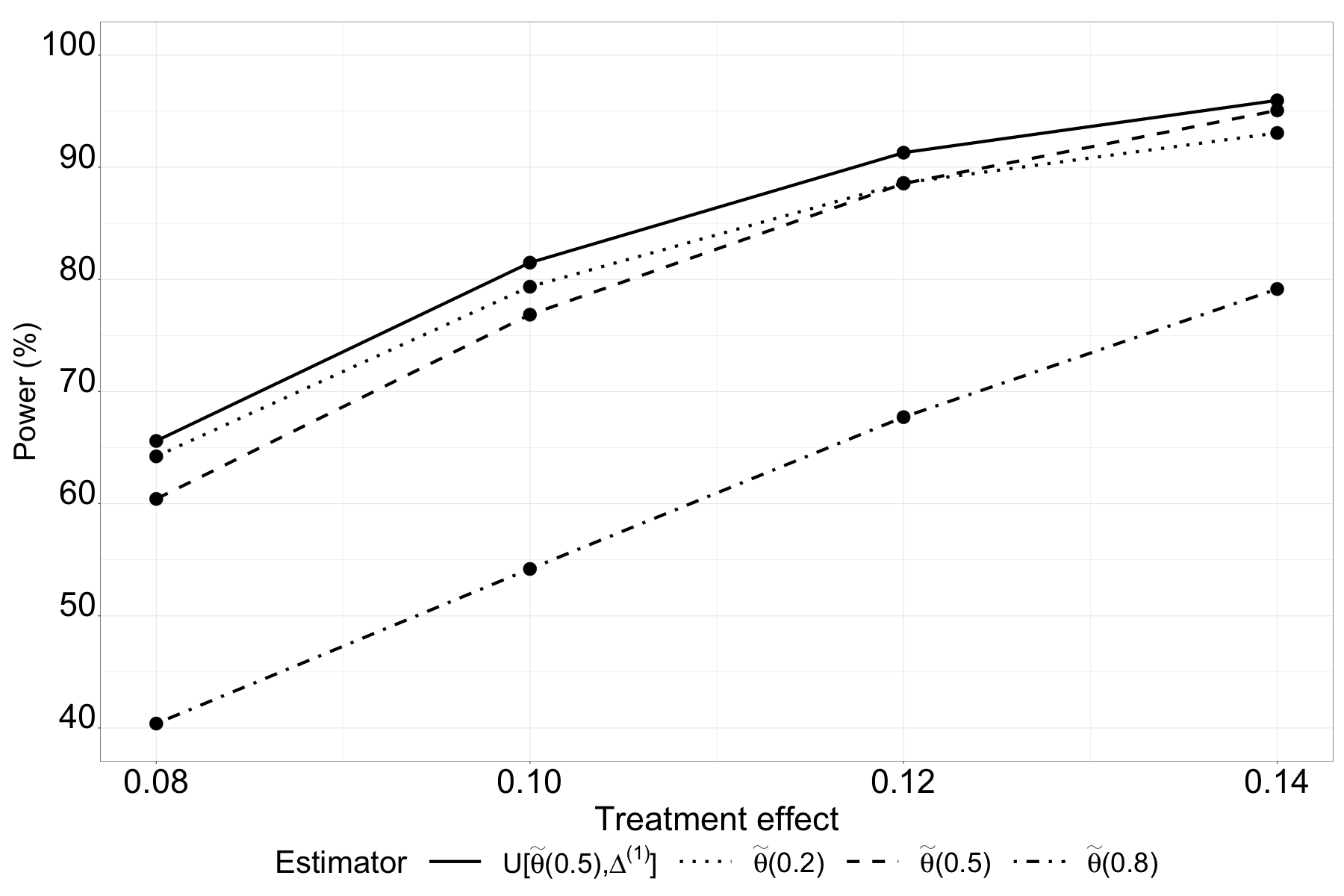}
	\caption{Consistently higher power of $U\left\{ \widetilde{\theta}(0.5), \Delta^{(1)} \right\}$ than $\widehat{\theta}(0.2)$, $\widehat{\theta}(0.5)$ and $\widehat{\theta}(0.8)$ to detect a promising treatment effect $\theta$ in the ACTT on COVID-19.}
	\label{F:power}
\end{figure}



\section{Discussion}
\label{sec:discussion}
In this article, we develop our ensemble framework by combining two base estimators with a linear formulation in (\ref{def_ux_motivate}). {An alternative approach is to directly train DNN with base estimators as input to minimize MSE. As compared with this alternative, our proposed method has tractable MSE studied in Theorem} \ref{theorem_3_bias} {and conditions of achieving MSE reduction in Corollary} \ref{cor:improv}. {Moreover, our method has a small bias when combining two unbiased base estimators, with the upper bound studied in Theorem 2 in the Supplemental Materials and simulation results in Table} \ref{sim:table_case}. {In addition to a reduced MSE, a smaller bias is also important for point estimation to more accurately estimate the parameter of interest, especially for the Adaptive COVID-19 Treatment Trial in Section} \ref{sec:adap}. 

{Our proposed framework is general, in the sense that DNN can be replaced by other prediction models, for example, Support Vector Machine (SVM) or Random Forest (RF). The approximation error in Proposition 2 and the estimation error in} (\ref{thm_2_rate_2_general}) {can be updated by corresponding results. As shown in Supplemental Table 4, DNN has a better performance than SVM and RF when estimating quantiles of scale-uniform distribution. In practice, one can implement our framework with different models and select the best one for a particular problem.} With modified combination weight in (\ref{w_opt}) to accommodate different objectives, our automatic framework can be widely applied to many statistical problems, for example, variable selection with high dimensional data. Even in scenarios with limited improvement, the ensemble estimator is still meaningful in confirming that the base estimators have satisfactory performance. 

There are some potential limitations of our method. The deep learning based approach requires additional training and computational time to obtain the estimator. However, as illustrated in our shared code, one can instantly compute the weight parameter based on well-trained neural networks and construct the ensemble estimator. Our method is currently under a parametric framework, such that the sampling distribution is known in Figure \ref{F:alg}. Future works include generalization to nonparametric settings, and making statistical inference of the parameter of interest based on the proposed estimator. 

\section*{Supplemental Materials}
Supplementary Materials including Appendices, Tables and Figures referenced in this article are available online. The R code and a help file to replicate results in the main article are available at \url{https://github.com/tian-yu-zhan/DNN_Point_Estimation}. 

\section*{Acknowledgements}
The authors thank the editorial board and reviewers for their constructive comments. 

This manuscript was supported by AbbVie Inc. AbbVie participated in the review and approval of the content. Tianyu Zhan is employed by AbbVie Inc., Haoda Fu is employed by Eli Lilly and Company, and Jian Kang is Professor in the Department of Biostatistics at the University of Michigan, Ann Arbor. Kang’s research was partially supported by NIH R01 GM124061 and R01 MH105561. All authors may own AbbVie stock.

\section*{Conflict of Interest}
No potential competing interest was reported by the authors.

\bigskip

\bibliographystyle{Chicago}

\bibliography{DNN_theory_ref}
\end{document}